\RequirePackage{fix-cm}
\documentclass[smallextended,natbib]{class/svjour3}

\journalname{Journal Name}
\usepackage[T1]{fontenc}
\usepackage{float}
\usepackage{amsmath,amssymb}
\usepackage{graphicx}
\graphicspath{{figures/}}
\usepackage[short]{optidef}
\usepackage{xcolor}
\usepackage{hyperref}
\hypersetup{colorlinks,urlcolor=blue,citecolor=.,linkcolor=.}
\usepackage{siunitx}
\usepackage[thinc]{esdiff}
\usepackage{multirow}

\usepackage{silence}
\usepackage{subcaption}

\renewcommand{\vec}[1]{\mathbf{#1}}

\usepackage{tikz}

\newcommand\submittedtext{%
  \footnotesize This preprint has not undergone peer review or any post-submission improvements or corrections. The Version of Record of this article is published in Optimization and Engineering, and is available online at https://doi.org/10.1007/s11081-025-10073-2.}

\newcommand\submittednotice{%
\begin{tikzpicture}[remember picture,overlay]
\node[anchor=south,xshift=-2cm,yshift=4cm] at (current page.south) {\fbox{\parbox{\dimexpr\textwidth-\fboxsep-\fboxrule\relax}{\submittedtext}}};
\end{tikzpicture}%
}

\begin{document}

\title{Energy-efficient torque allocation for straight-line driving of electric vehicles based on pseudoconvex polynomials
}

\titlerunning{Energy-efficient torque allocation for EVs based on pseudoconvex polynomials}        % if too long for running head

\author{Josip Kir Hromatko \and \v{S}andor Ile\v{s} \and Branimir \v{S}kugor \and Jo\v{s}ko Deur}
\institute{
    Josip Kir Hromatko \and \v{S}andor Ile\v{s} \\
    Faculty of Electrical Engineering and Computing,  University of Zagreb, Croatia \\
    \email{\{josip.kir.hromatko, sandor.iles\}@fer.unizg.hr} \\
    \vspace{.1cm}\\
    Branimir \v{S}kugor \and Jo\v{s}ko Deur \\
    Faculty of Mechanical Engineering and Naval Architecture, University of Zagreb, Croatia \\
    \email{\{branimir.skugor, josko.deur\}@fsb.unizg.hr}\\
    \vspace{.1cm}\\
    Corresponding author: Josip Kir Hromatko
}

\date{Received: date / Accepted: date}
% The correct dates will be entered by the editor

\maketitle
\submittednotice

\begin{abstract}
Electric vehicles with multiple motors provide a flexibility in meeting the driver torque demand, which calls for minimizing the battery energy consumption through torque allocation. In this paper, we present an approach to this problem based on approximating electric motor losses using higher-order polynomials with specific properties. To ensure a well-behaved optimization landscape, monotonicity and positivity constraints are imposed on the polynomial models using sum of squares programming. This methodology provides robustness against noisy or sparse data, while retaining the computational efficiency of a polynomial function approximation. The torque allocation problem based on such polynomials is formulated as a constrained nonlinear optimization problem and solved efficiently using readily available solvers. In the nominal case, the first-order necessary conditions for optimality can also be used to obtain a global solution. The performance of the proposed method is evaluated on several certification driving cycles against a grid search-based benchmark. Results show a modest influence on electric energy consumption, while enabling real-time optimization and integration with other vehicle control systems.
\keywords{Electric vehicle control \and Torque allocation \and Nonlinear optimization \and Pseudoconvexity \and Polynomial regression \and Sum of squares}  % 4 to 6
% \PACS{PACS code1 \and PACS code2 \and more}
% \subclass{MSC code1 \and MSC code2 \and more}
\end{abstract}

\clearpage
\section{Introduction}
\label{sec:intro}

\subsection{Background and problem formulation}

The increasing usage of electric vehicles (EVs) in everyday life has enabled improvements in both passenger safety and vehicle energy efficiency. To allow for a smoother transition from conventional vehicles to EVs, it is important that the supporting infrastructure (i.e., charging stations) meets the demand. In parallel, significant engineering efforts are put into extending the driving range of electric vehicles, which is often the deciding factor for potential buyers.

One way to extend the driving range is to incorporate powertrain efficiency information into vehicle control systems such as the torque allocation module~\citep{eff_tv_survey}. Several approaches to this problem rely on offline optimization, producing a set of simple-to-implement rules or lookup tables~\citep{lenzo2017torque,koehler2017}. However, generating such rules under varying conditions could become ineffective as it would require repeating the procedure for each specific set of vehicle or environment parameters. Additionally, rule-based methods cannot provide instant adaptability to vehicle parameter changes or constraints imposed by other vehicle control systems (e.g., varying torque limits \citep{prost2024energy}, desired driver comfort \citep{ou2020} or braking regulations \citep{xu2020}). 

When using online optimization-based methods (such as model predictive control) for torque allocation, it is beneficial to model powertrain losses in a way that facilitates efficient optimization~\citep{eff_tv_nmpc}. In general, numerical issues and suboptimal solutions due to local minima can be avoided by enforcing convexity of the optimization problem \citep{dizqah2020}. On the other hand, such a relaxation sometimes leads to large modeling errors and performance loss. An approach that allows for nonlinearities, but still retains some desirable properties is to enforce pseudoconvexity \citep{mangasarian} of the motor loss function approximation, which decreases the modeling error while retaining computational efficiency. Furthermore, when choosing a class of admissible functions for regression, it is desirable that the approximation accuracy is retained also with noisy or sparse data. While higher-order polynomials and more powerful function approximators such as neural networks provide greater accuracy, there is a risk of overfitting and producing unrealistic behavior, which reduces the generalization capabilities of the method.

\begin{figure}
    \centering
    \includegraphics[width=.5\linewidth]{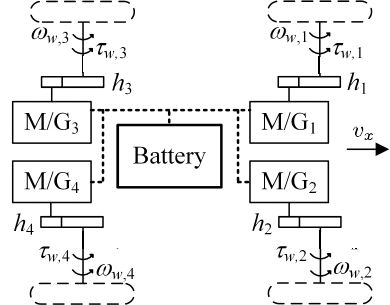}
    \caption{The powertrain configuration considered in this paper}
    \label{fig:powertrain}
\end{figure}

In this paper, we address the problem of optimizing torque allocation for minimal energy consumption of electric vehicles during straight-line driving. It is assumed that the vehicle has four independent fixed-gear electric motors, with equal or unequal motors driving the wheels on the same axle, as shown in Figure~\ref{fig:powertrain}. The motor loss data are modeled using pseudoconvex polynomials, providing a compromise between approximation accuracy and computational efficiency. Finally, the approach is directly applicable to the special case of 4WD powertrain, having two motors on front and rear axles, connected to the wheels via open differentials.

\subsection{Related literature}

The authors of \citep{dizqah2016} and \citep{filippis2018} show that electric motor losses can be approximated using a monotonically increasing cubic polynomial with a single inflection point. For the powertrain with equal motors, the former reference derives a simple and optimal rule-based strategy for straight-line driving, involving low-torque single-axle and high-torque equal distribution operation, with the switching torque defined as a function of vehicle velocity. The latter reference derives constraints on the cubic polynomial coefficients such that the fitted polynomial is guaranteed to satisfy the specific properties of the motor loss function. Section 4.2 and Figure 7 in a related paper \citep{lenzo2017torque} show that a higher fitting accuracy can be obtained by using spline functions, i.e., a more flexible function class. However, to the best of our knowledge, similar explicit constraints for higher-order polynomials have not been established. Additionally, experimental data in \citep{dizqah2016} show that the motor loss function does not necessarily have only one inflection point.

Several papers propose methods for energy-efficient torque allocation for electric vehicles with different motors. In \citep{lenzo2017torque}, the authors expand on \citep{dizqah2016} and derive analytic expressions for such powertrains based on loss approximation by cubic polynomials. It is assumed that the powertrain configuration with different motors uses the same motor technology, but different torque range on the two axles. A related paper \citep{dizqah2020} proposes a method for the more general case based on Taylor series expansion, similar to the SQP method \citep{biegler2010nonlinear} and leading to a (parametric) quadratic optimization problem. Both papers suggest an implementation in the form of lookup tables, with the goal of minimal computational demands.

Online torque distribution algorithms were considered in several papers as well. In \citep{yang2017research}, the authors compare several optimization methods (some of which could run in real-time), while others propose more generic optimization methods such as genetic algorithms \citep{xu2020} and particle swarm optimization \citep{ou2020}. However, such methods might be too computationally demanding to run on automotive-grade hardware. An alternative approach to online optimization, as proposed in \citep{chen2011fast}, is leveraging the Karush-Kuhn-Tucker (KKT) conditions to efficiently identify all potential optimal solutions for certain types of efficiency maps, and subsequently discarding the suboptimal ones.

An approach similar to ours was proposed in \citep{menner2021}, where the authors incorporated pseudoconvexity constraints into kernel regression, a highly flexible function approximation method. However, kernel regression's flexibility comes at the cost of requiring a large number of parameters and higher computational demands compared to higher-order polynomials. Furthermore, the implementation details of enforcing the pseudoconvexity constraints on kernel regression were not reported.

Higher-order polynomial approximations offer a compelling alternative, potentially achieving comparable accuracy with fewer parameters and a lower computational cost. On the other hand, a careful consideration of their approximation accuracy is needed, especially regarding desirable properties such as positivity and monotonicity. Fitting polynomial functions with such properties on a specific interval can be achieved using methods based on sums of squares \citep{magnani2005tractable}. These methods are commonly used in applications related to operations research or quantum theory, but also for proving stability of dynamical systems or robust control system design \citep{parrilo2012}.

\subsection{Statement of contributions}
In this paper, we present a computationally efficient method which can be used to approximate motor losses accurately, while retaining real-time capabilities for further integration with other vehicle control systems. The main contributions are:
\begin{itemize}
    \item A method for approximating electric motor losses with a positive and monotonically increasing polynomial of arbitrary order, based on a sum of squares formulation.
    \item A real-time feasible, gradient-based torque allocation strategy that minimizes energy consumption of an electric vehicle using the polynomial loss approximation.
    \item A comparison with a method for identifying candidate solutions using first-order necessary (KKT) conditions.
\end{itemize}

\subsection{Organization}
The remainder of this paper is
structured as follows: Section \ref{sec:model} presents the mathematical model of vehicle dynamics and powertrain losses. Section \ref{sec:fit} presents the method of fitting pseudoconvex polynomials based on sums of squares. Section \ref{sec:opti} describes the torque allocation strategies for two possible powertrain configurations. Section \ref{sec:results} presents the approximation accuracy of the developed method and simulation results for energy usage on several standard driving cycles. Final comments are given in Section \ref{sec:outro}.

\section{Vehicle modelling}
\label{sec:model}
This section presents the vehicle dynamics and powertrain model used in the experiments, based on the one described in \citep{skugor2024optimization}. Additionally, key simplifying assumptions are stated.

\subsection{Longitudinal vehicle dynamics}
The longitudinal motion of the vehicle can be described by:
\begin{equation}
    m\dot{v}_x=F_x - R_\mathrm{r}\cdot m\cdot g - 0.5\cdot\rho_\mathrm{air}\cdot C_\mathrm{d}\cdot A_\mathrm{f}\cdot v_x^2
\end{equation}
where $m$ denotes the vehicle mass, $v_x$ the longitudinal speed, $F_x$ the total traction force, $R_\mathrm{r}$ the rolling resistance coefficient, $g$ the gravitational constant, $\rho_\mathrm{air}$ the air density, $C_\mathrm{d}$ the air resistance coefficient and $A_\mathrm{f}$ the frontal area. The vehicle is assumed to be driving on a flat surface, neglecting road slope effects.

The traction force is generated through the four tires:
\begin{equation}
    F_x = \sum_{i=1}^4 F_{\mathrm{w},i}
\end{equation}
where we assume no tire slip and a direct relation between the tire forces $F_{\mathrm{w},i}$ and wheel torque $\tau_{\mathrm{w},i}$:
\begin{equation}
    F_{\mathrm{w},i} = \frac{\tau_{\mathrm{w},i}}{R},\qquad i=1,\dots,4
\end{equation}
where $R$ denotes the effective wheel radius.

\subsection{Powertrain model}
In this paper, a four-motor EV configuration is considered, where each motor drives one of the wheels. The motor torques $\tau_{\mathrm{m},i}$ and speeds $\omega_{\mathrm{m},i}$ are related to wheel torques $\tau_{\mathrm{w},i}$ and speeds $\omega_{\mathrm{w},i}$ through fixed gear ratios $h_i$:
\begin{equation}\label{eq:gear}
    \omega_{\mathrm{m},i} = h_i\cdot\omega_{\mathrm{w},i},\quad \tau_{\mathrm{m},i} = \frac{1}{h_i}\cdot\tau_{\mathrm{w},i},\quad i=1,\dots,4
\end{equation}
Additionally, an ideal transmission is assumed, along with symmetrical loss maps for braking and acceleration. The electric power of a single motor can be calculated as:
\begin{equation}
    P_{\mathrm{el},i} = \omega_{\mathrm{m},i}\cdot\tau_{\mathrm{m},i} + P_{\mathrm{m,loss},i}(\omega_{\mathrm{m},i},\tau_{\mathrm{m},i})
\end{equation}
where the power loss term $P_{\mathrm{m,loss},i}$ is determined by interpolating the motor loss map at a given operating point. The total energy consumption over a driving cycle of duration $t_\mathrm{f}$ is obtained by integrating the electric power of the four motors:
\begin{equation}
    E_\mathrm{el,t} = \int_{t=0}^{t_\mathrm{f}} \left(\sum_{i=1}^4 P_{\mathrm{el},i}(t)\right)dt
\end{equation}

\section{Pseudoconvex polynomial regression}
\label{sec:fit}
This section introduces the method of approximating the motor losses by pseudoconvex polynomials. It extends the proposition that power losses increase monotonically with torque demand \citep{dizqah2016} from third-order to higher-order polynomials. Since symmetrical power loss maps are assumed, it is sufficient to consider only nonnegative motor torques. The objective here is to fit a function $f$ to a given dataset using a least-squares fit:
\begin{equation}
    \min_{f\in\mathcal{F}} \sum_{i=1}^N (y_i-f(x_i))^2
\end{equation}
where $x_i$ and $y_i$ denote the input and output data, $N$ is the total number of samples and $\mathcal{F}$ represents the class of admissible functions (e.g., polynomial functions).

\subsection{Pseudoconvex functions}
Generally, a function $f\colon \mathbf{R}^n\rightarrow\mathbf{R}$ is called \textit{quasiconvex} \citep{boyd2004convex} if its domain and all its sublevel sets:
\begin{equation}
    S_\alpha = \left\{\vec{x}\in\text{dom} f\ |\ f(\vec{x})\leq\alpha\right\},\quad \alpha\in\mathbf{R}
\end{equation}
are convex. Furthermore, a differentiable quasiconvex function is \textit{pseudoconvex} \citep{mangasarian} if and only if its domain is convex and for all ($\vec{x}$,$\vec{y}$) from its domain the following holds:
\begin{equation}\label{eq:pk}
    f(\vec{y})\leq f(\vec{x})\ \Rightarrow\ \nabla f(\vec{x})^\top\cdot (\vec{y}-\vec{x})\leq 0
\end{equation}
Finally, a scalar function $f\colon \mathbf{R}\rightarrow\mathbf{R}$ is pseudoconvex if and only if it is monotonically increasing for $x>x^*$ and monotonically decreasing for $x<x^*$, where $x, x^* \in \text{dom } f$ and $x^*$ is a global minimizer of $f$.

\subsection{Sum of squares decomposition}
In this section, we consider electric motor losses at a fixed speed $\omega_\mathrm{m}$, i.e., a scalar motor loss function $P_\mathrm{m,loss}(\tau_\mathrm{m})$. As stated in \citep{dizqah2016} and supported by experimental data, this function is monotonically increasing with the motor torque. For such a function, the pseudoconvexity condition \eqref{eq:pk} is satisfied by imposing the following condition on its derivative:
\begin{equation}\label{eq:der_constr}
    \diff{f}{x}\geq 0,\quad\forall x\in\mathcal{X}
\end{equation}
where $\mathcal{X}$ denotes the range of interest (e.g., the torque range of the motor).

However, imposing the constraint \eqref{eq:der_constr} for a polynomial function of arbitrary order is challenging \citep{curmei2023shape}. A standard relaxation method is to replace the nonnegativity constraint by a constraint that involves sum of squares polynomials. Although the following is restricted to the univariate case, the statements hold for multivariate polynomials as well.

Generally, a univariate polynomial $p(x)$ of order $2d$ is a sum of squares (sos) if there exist polynomials $q_1,\dots,q_m$ of order $d$ such that:
\begin{equation}
    p(x)=\sum_{k=1}^m q_k^2(x)
\end{equation}
For univariate polynomials, being a sum of squares is both a necessary and a sufficient condition for nonnegativity \citep{Hilbert1888}. Furthermore, a univariate polynomial $p(x)$ of order $2d$ is nonnegative on the interval $[0,\infty)$ if and only if it can be written as:
\begin{equation}
    p(x)=t(x)+x\cdot s(x)
\end{equation}
where $s(x)$ and $t(x)$ are sums of squares. If $\deg(p) = 2d$, then $\deg(t) \leq 2d$ and $\deg(s) \leq 2d-2$, while if $\deg(p) = 2d+1$, then $\deg(t) \leq 2d$ and $\deg(s) \leq 2d$.

\subsection{Constrained least-squares fitting}
The objective of fitting a polynomial $p(x)$ to the dataset ($x_i,y_i$), $i=1,\dots,N$, while ensuring that $p(x)$ is a positive and monotonically increasing function for nonnegative $x$, can be formulated as an sos program:
\begin{mini!}
    {z}{\sum_{k=1}^N \left(y_i-p(x_i)\right)^2}{\label{eq:opt_sos}}{\label{eq:opt_sos_cost}}
    \addConstraint{
    p(0)>0}
    \addConstraint{
    p'(x)-x\cdot s(x)\quad\text{is sos}}
    \addConstraint{
    s(x)\quad\text{is sos}\label{eq:sos_constr}}
\end{mini!}
where $z$ denotes a vector containing the coefficients of $p(x)$ and $s(x)$. Also, $p'(x)$ represents the derivative of $p(x)$ with respect to $x$. If strict positivity of the derivative is required (e.g., to avoid local plateaus), the condition \eqref{eq:sos_constr} can be modified as:
\begin{equation}\label{eq:derpos}
    s(x)-\epsilon\quad\text{is sos}
\end{equation}
for some $\epsilon>0$.
The resulting sum of squares program can be reformulated as a (potentially larger) semidefinite program (SDP) and solved efficiently with modern solvers.

\subsection{Relation to analytic conditions for univariate cubics}
The authors of \citep{lenzo2017torque} state that a cubic polynomial:
\begin{equation}
    p(x)=a x^3+b x^2+c x+d
\end{equation}
is monotonically increasing if $a>0$, $c>0$ and $b^2<3 a c$. In this section we show that these conditions are equivalent to requiring that $p'(x)$ is a sum of squares.

The derivative of the given polynomial is:
\begin{equation}
    p'(x) = 3 a x^2 + 2 b x + c
\end{equation}
which can be rewritten as:
\begin{equation}\label{eq:pder}
    p'(x) = 3 a\cdot \left(x^2+\frac{2 b}{3 a} x + \frac{c}{3 a}\right) = 3 a\cdot  q(x)
\end{equation}
The polynomial $q(x)$ can be rewritten as:
\begin{equation}
    q(x) = x^2+\frac{2 b}{3 a} x + \frac{c}{3 a} = \left(x+\frac{b}{3 a}\right)^2 + e^2
\end{equation}
where
\begin{equation}\label{eq:esqrt}
    e = \sqrt{\frac{3ac-b^2}{9a^2}}
\end{equation}
The polynomial $q(x)$ is a sum of squares if the radicand in \eqref{eq:esqrt} is positive, i.e., if $b^2<3ac$. Then, from \eqref{eq:pder} it follows that $p'(x)$ is a sum of squares if $a>0$. Finally, these two conditions imply that $c>0$ must be true as well.

\subsection{Fitting multivariate polynomials}
For multivariate polynomials, nonnegativity generally does not imply that a polynomial is a sum of squares (except for bivariate quartics). However, the sum of squares constraints presented so far can easily be stated in the multivariate case, ensuring positive and monotonically increasing multivariate polynomials. A potential use case could be to fit the motor losses jointly depending on the motor torque and speed, i.e., to consider $P_\mathrm{m,loss}(\omega_\mathrm{m},\tau_\mathrm{m})$. While such an approach is not considered in this paper, it will be the focus of future research.

\subsection{Practical considerations}
In practice, fitting pseudoconvex polynomials has both advantages and drawbacks. On one hand, such a method guarantees desired properties of the polynomial, while ensuring robustness to noise or sparse data. However, fitting polynomials of higher order (above 10, based on our current experience) can lead to numerical issues, infeasibilities and long optimization run times. Fortunately, due to their presence in various scientific fields, a steady improvement in software related to solving sos programs can be expected.
\section{Torque allocation strategy}
\label{sec:opti}
The problem of minimizing energy usage through torque allocation can be formulated as an optimization problem, which we present here for two possible powertrain configurations: (i) with all four motors being equal, and (ii) with front motors differing from rear motors. As shown in~\citep{dizqah2016} and~\citep{skugor2024optimization}, the optimal left-right split for straight driving is to apply equal torques to both sides. This suggests that it is sufficient to consider only one side of the vehicle (i.e., the front-rear torque distribution). Also, no wheel slip is assumed, leading to equal rotational speeds for all wheels. Additionally, the front-to-total wheel torque distribution ratio is defined as:
\begin{equation}
    \sigma = \frac{\tau_\mathrm{w,f}}{\tau_\mathrm{w,f} + \tau_\mathrm{w,r}}
\end{equation}
which relates the front and rear wheel torques $\tau_\mathrm{w,f}$, $\tau_\mathrm{w,r}$ with the total torque demand $\tau_\mathrm{w,ref}$:
\begin{align}
    \tau_\mathrm{w,f} &= \sigma\cdot\tau_\mathrm{w,ref}\\
    \tau_\mathrm{w,r} &= (1-\sigma)\cdot\tau_\mathrm{w,ref}
\end{align}
The corresponding motor torques $\tau_\mathrm{m,f}$ and $\tau_\mathrm{m,r}$ are proportional to the wheel torques as given in \eqref{eq:gear}.

\subsection{Equal motors}
If all four motors are assumed to be identical, it is possible to formulate the optimization problem in terms of the distribution ratio and the torque demand:
\begin{mini!}
    {\sigma}{P_{\mathrm{m,loss}}(\sigma\cdot\tau_\mathrm{w,ref}) + P_{\mathrm{m,loss}}((1-\sigma)\cdot\tau_\mathrm{w,ref})}{\label{eq:opt_split_1}}{\label{eq:opt_split_1_cost}}
    \addConstraint{
    0.5 \leq \sigma \leq 1}
\end{mini!}
where $P_{\mathrm{m,loss}}(\tau)$ denotes the loss model at a given motor speed. Since the objective function \eqref{eq:opt_split_1_cost} is symmetric with respect to $\sigma$, it is sufficient to consider the distribution ratios only in the range [0.5,1]. The final motor torques can be determined based on factors such as vehicle acceleration (e.g., allocating more torque to the front wheels during braking) or safety considerations (e.g., using front motors to induce understeer rather than oversteer in low-traction conditions).

\subsection{Unequal motors}
If the front and rear motors are different, some modifications of the torque allocation problem are needed:
\begin{mini!}
    {\sigma}{P_{\mathrm{m,loss,f}}(\sigma\cdot\tau_\mathrm{w,ref}) + P_{\mathrm{m,loss,r}}((1-\sigma)\cdot\tau_\mathrm{w,ref})}{\label{eq:opt_split_2}}{\label{eq:opt_split_2_cost}}
    \addConstraint{
    0 \leq \sigma \leq 1}
\end{mini!}
where $P_{\mathrm{m,loss,f}}$ and $P_{\mathrm{m,loss,r}}$ are generally distinct functions of motor losses at motor speeds $\omega_\mathrm{m,f}$ and $\omega_\mathrm{m,r}$. The motor speeds are related to the wheel speeds through the fixed gear ratios $h_i$ as given in \eqref{eq:gear}. In this case, the optimal torque distribution ratio can vary between 0 and 1 since the objective function \eqref{eq:opt_split_2_cost} is not necessarily symmetric with respect to $\sigma$.

\subsection{Discussion}

When approximating motor losses using a cubic polynomial with pseudoconvexity constraints, the resulting function has a single inflection point. For identical motors, the torque allocation objective function is either convex or concave, depending on the torque request. If the function is convex, the optimal solution distributes torque equally among the motors; if concave, the solution favors single-axle operation. This can be extended to unequal motors by assuming a relation between the motor loss functions (e.g., through scaling). However, when applied to general unequal motors, this approximation can result in multiple local optima. Furthermore, the cubic polynomial has a larger approximation error compared to higher-order polynomials.

To reduce the chance of suboptimal solutions, global optimization methods such as genetic algorithms or particle swarm optimization can be used. However, most of these approaches are not real-time feasible. Another approach would be to use the Karush-Kuhn-Tucker (KKT) conditions to eliminate suboptimal solutions, as suggested in \citep{chen2011fast}. Since the cost function in the control allocation problem is represented by a polynomial, its gradient can be easily computed. Identifying potential optimal solutions involves finding the roots of this polynomial and comparing the corresponding cost values with those at the constraint boundaries. For more general optimization problems, the KKT-based solution can serve as an initial condition for solving the nonlinear optimization problem.

Additionally, the torque allocation problem can be easily reformulated directly in terms of motor torques (instead of the distribution ratio and torque request). This would facilitate including additional constraints or objectives related to varying torque limits, reducing control input oscillations or meeting braking regulations.

\section{Results}
\label{sec:results}
\begin{figure*}
\centering
\begin{subfigure}{0.49\textwidth}
    \includegraphics[width=\textwidth]{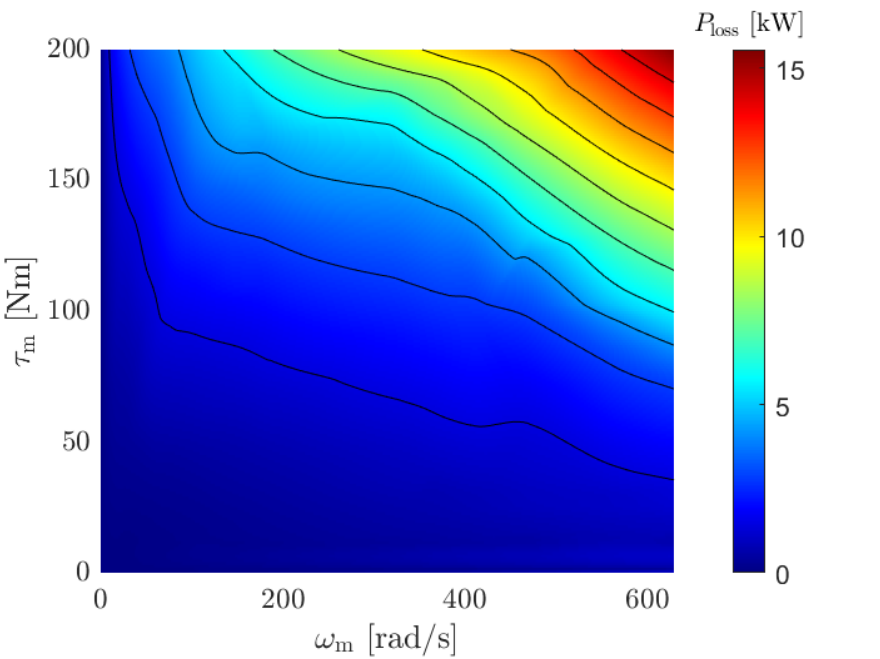}
    \caption{Motor type 1}
    \label{fig:loss_map_1}
\end{subfigure}
\hfill
\begin{subfigure}{0.49\textwidth}
    \includegraphics[width=\textwidth]{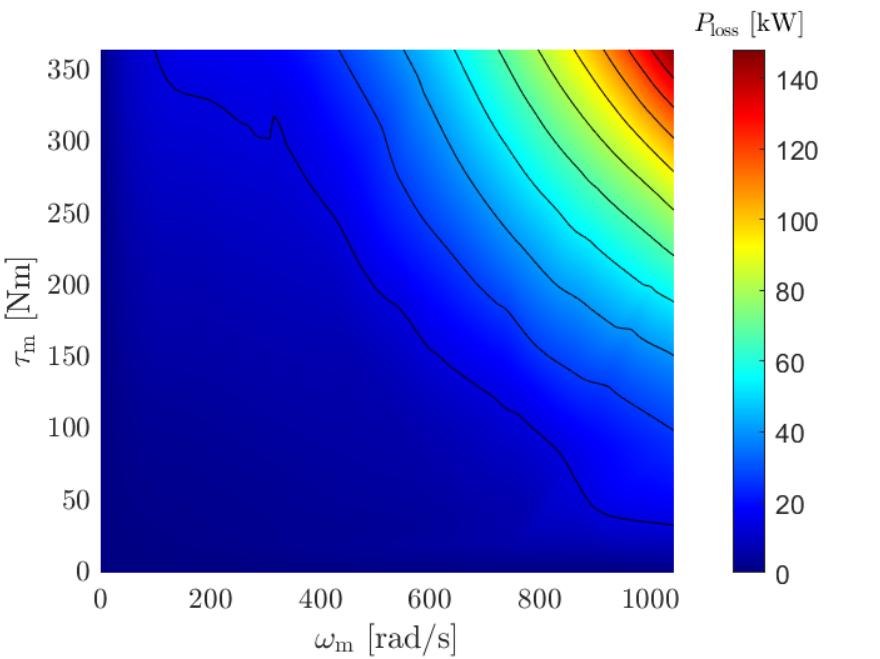}
    \caption{Motor type 2}
    \label{fig:loss_map_2}
\end{subfigure}
\caption{Electric loss maps for the two considered motors}
\label{fig:loss_map}
\end{figure*}

In this section, we consider a specific electric vehicle and present the power loss fitting accuracy using the pseudoconvex polynomial approach. We also investigate the optimal torque distribution ratios and electric energy consumption on several certification driving cycles. The vehicle parameters, given in Table~\ref{tab:params}, correspond to a modified Chevrolet Volt extended range electric vehicle (EREV)~\citep{deur2012modeling}. The power loss maps for the two electric motors, shown in Figure~\ref{fig:loss_map_1} and Figure~\ref{fig:loss_map_2}, are derived from the motors' efficiency maps originally provided in \citep{grebe2011} and \citep{miller2011gm}.

\begin{table}
    \centering
    \caption{Vehicle parameters}
    \label{tab:params}
    \begin{tabular}{c c c c}
         \hline\noalign{\smallskip}
         Parameter & Value & Unit & Description\\
         \noalign{\smallskip}\hline\noalign{\smallskip}
          $m$ &   2200  & kg & mass of the vehicle\\
          $R$ &  0.317  & m & effective wheel radius\\
          $R_\mathrm{r}$ & 0.008 & - & rolling resistance coefficient \\
          $C_\mathrm{d}$ & 0.275 & - & air resistance coefficient \\
          $A_\mathrm{f}$ & 2.22 & m$^2$ & frontal area \\
          $\rho_\mathrm{air}$ & 1.225 & kg$\cdot$m$^{-3}$ & air density \\
          $g$ & 9.81 & m$\cdot$s$^{-2}$ & gravitational constant \\
          $h_1$ & 4.4863 & - & gear ratio for motor type 1 \\
          $h_2$ & 7.425 & - & gear ratio for motor type 2 \\
          \noalign{\smallskip}\hline
    \end{tabular}
\end{table}

\subsection{Simulation framework}
In the experiments, we made a few assumptions. Firstly, the vehicle losses were simplified and include only the motor electric losses, ignoring additional sources such as tire or friction losses. This was motivated by available literature~\citep{skugor2024optimization, lenzo2017torque} showing that the electric losses are the dominant factor in total vehicle losses, while secondary factors have a negligible effect. Also, this enabled a better comparison between different methods, since the same amount of information is consistently used in the optimization. Secondly, it was assumed that the motor maps are identical for negative torques, i.e., there is no difference (from the optimization standpoint) between traction and braking. Extending the method presented in this paper to asymmetric loss maps will be the topic of future work. Finally, drivetrain dynamics were neglected for simplicity.

The following results were obtained using MATLAB 2021b on a Windows 10 machine with 8 GB of RAM and an Intel Core i5-10300H processor running at 2.50 GHz.

\subsection{Motor loss fitting}
The constrained least-squares fitting problems of the form given by \eqref{eq:opt_sos} were implemented using the open-source toolbox YALMIP~\citep{yalmip} and its sum of squares module~\citep{yalmip_sos}. The underlying SDPs were solved using MOSEK~\citep{mosek}. Also, the derivative positivity constant in \eqref{eq:derpos} was chosen as $\epsilon=10^{-3}$.

The fit was performed using polynomials of order $d$ (specified in each section), for each of the $n_\omega$ speeds in the motor loss maps (with a step of 1 rad/s), leading to $n_\omega\times (d+1)$ polynomial coefficients. The results were compared with an unconstrained least-squares fit obtained using the MATLAB command \texttt{polyfit}.

\subsubsection{Our data}
\begin{figure}
    \centering
    \includegraphics[width=.6\linewidth]{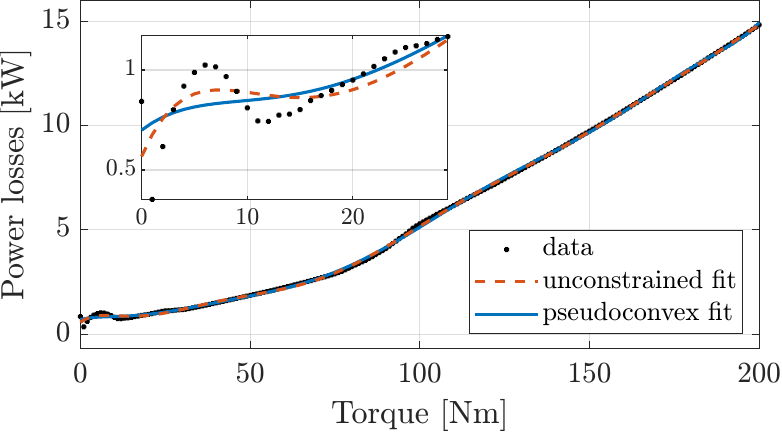}
    \caption{Polynomial fits of 10th degree, using loss map 1 at $\omega_\mathrm{m}=300$ rad/s}
    \label{fig:fit_full}
\end{figure}

Figure~\ref{fig:fit_full} shows examples of polynomial fits, one without any constraint and another with the presented pseudoconvex constraints. Generally, there is no visible difference between the two approximation methods. However, at low torques (zoomed in part of the figure), power losses locally exhibit oscillatory behavior. This is caused by a linear interpolation of the originally low-resolution efficiency map (see~\citep{miller2011gm}) used for calculating the power losses presented in \citep{deur2012modeling}. From the perspective of the whole power loss range, the discrepancy can be considered negligible. In this region, the fitted unconstrained polynomial does not retain monotonicity. While the effect can be avoided by modifying the loss data locally, the pseudoconvex fit provides the desired properties, allowing for a better agreement with our assumptions about electric motor losses.

\subsubsection{Data from the literature}
\begin{figure*}
\centering
\begin{subfigure}{0.49\textwidth}
    \includegraphics[width=\linewidth]{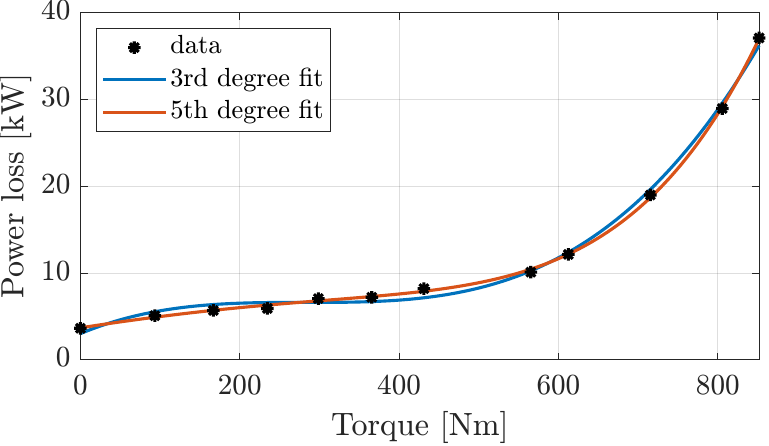}
    \caption{3rd and 5th order pseudoconvex polynomials}
    \label{fig:fit_dizqah_3vs5}
\end{subfigure}
\hfill
\begin{subfigure}{0.49\textwidth}
    \includegraphics[width=\linewidth]{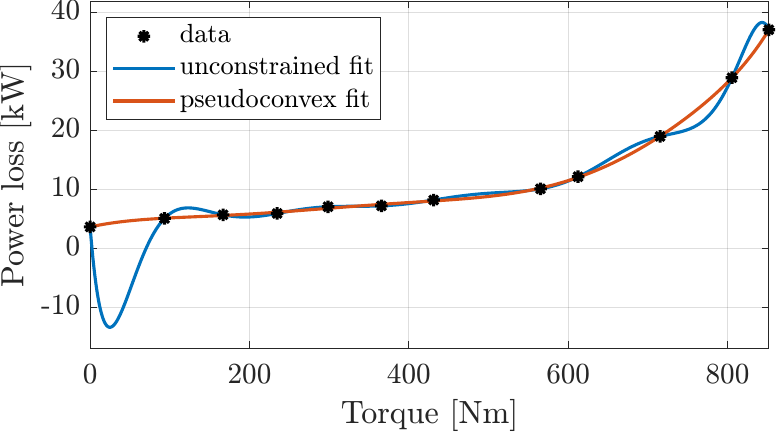}
    \caption{11th order unconstrained and pseudoconvex polynomials}
    \label{fig:fit_dizqah_11}
\end{subfigure}
\caption{Polynomial fitting results using data from Fig.~4 in \citep{dizqah2016}, at 105 km/h}
\label{fig:fit_dizqah}
\end{figure*}

If experimental data from~\citep{dizqah2016} are considered instead, the benefits of using higher-order polynomials are more easily seen. Figure~\ref{fig:fit_dizqah_3vs5} compares the results of using a 3rd order and a 5th order polynomial. The data suggest that imposing a single inflection point might be a limiting assumption. In both cases, pseudoconvexity constraints were included. As expected, a higher accuracy is obtained with the higher-order polynomial, while retaining the desired properties. 

On the other hand, as shown in Figure~\ref{fig:fit_dizqah_11}, using a high-order polynomial without additional constraints can diminish the approximation accuracy. Although the order of the polynomial is very high compared to the number of samples in this case (and not very likely to be used in practice), one can observe that the pseudoconvex approximation does not produce the same unwanted effects. This implies that the proposed method could be a viable option even with very few samples and without any additional preprocessing (i.e., increasing the map density through interpolation).

\subsection{Total powertrain losses}
In this section, we analyze the objective function of the torque allocation problem (i.e., total powertrain losses) for the two considered configurations. Individual motor losses were approximated using 10th degree pseudoconvex polynomials. Although the imposed property of monotonicity does not necessarily extend to the total powertrain losses, some conclusions can still be drawn.

\begin{figure*}
\centering
\begin{subfigure}{0.49\textwidth}
    \includegraphics[width=\linewidth]{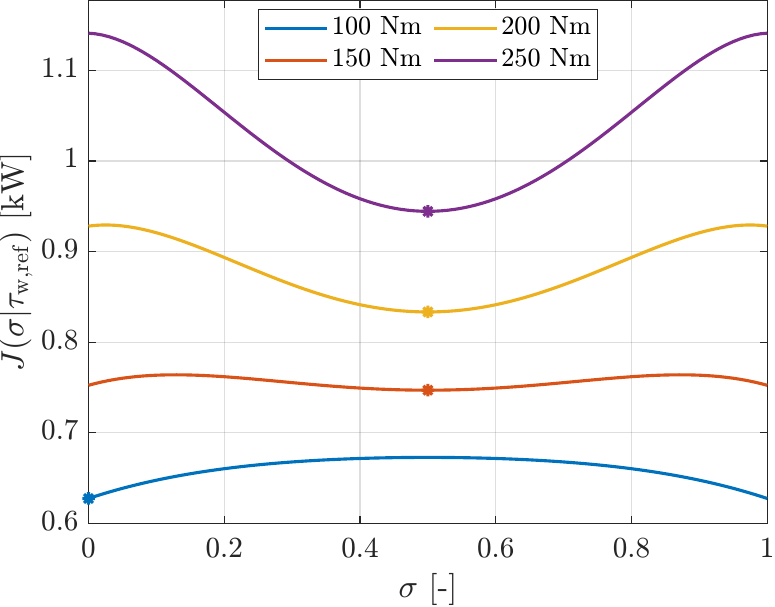}
    \caption{Equal motors}
    \label{fig:total_loss_1}
\end{subfigure}
\hfill
\begin{subfigure}{0.49\textwidth}
    \includegraphics[width=\linewidth]{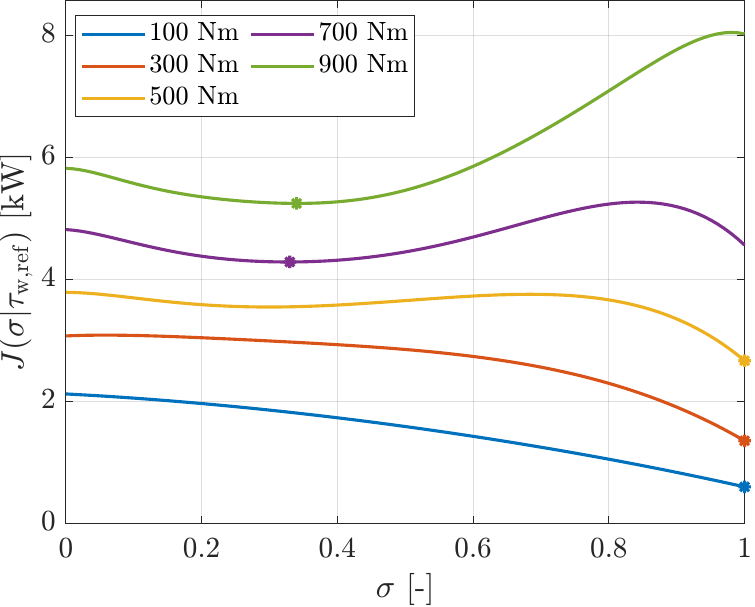}
    \caption{Unequal motors}
    \label{fig:total_loss_2}
\end{subfigure}
\caption{Total powertrain losses for varying torque demands at $\omega_\mathrm{w}=50$ rad/s; the optimal torque distribution is marked with a star}
\label{fig:total_loss}
\end{figure*}

Figures~\ref{fig:total_loss_1} and~\ref{fig:total_loss_2} show the total powertrain losses as a function of the torque distribution ratio $\sigma$ and the wheel torque demand for one side of the vehicle $\tau_\mathrm{w,ref}$. When unequal motors are considered, the front ones are set to motor type 1 (weaker) and the rear ones to motor type 2 (stronger).

As shown in~\citep{dizqah2016} and~\citep{skugor2024optimization}, the configuration with equal motors has a specific \textit{switching torque demand}, above which the optimal distribution is equal to 0.5, regardless of the motor characteristics. If the torque request is below that point, the optimal distribution ratio is either 0 or 1 (with the same cost due to the loss function symmetry).

However, when unequal motors are considered, the optimal torque allocation strategy is not as straightforward. For low torque demands, only the weaker motors should be used. As the torque demand increases, stronger motors should be engaged as well. The optimal torque distribution ratio will in general depend on the specific characteristics of the two motors and the total torque request.

From the presented figures, one could conclude that the objective function has at most three local minima. Proving such a property (perhaps together with imposing additional constraints on the individual motor losses) and finding a deterministic optimization method with a globally optimal solution will be investigated in future work.

\subsection{Optimal torque distribution}
In this section, we compare the optimal torque distribution ratios obtained by different optimization methods, for the two considered powertrain configurations. The wheel speed and torque demand ranges were determined based on the certification driving cycles presented in the next section. Besides the gradient-based strategies with unconstrained and pseudoconvex polynomial loss approximation, we include the control allocation approach described in~\citep{skugor2024vsd}. With this method, a grid-based search over the motor loss maps is used instead, which guarantees global optimality (up to the grid step length).

The torque allocation problem with polynomial loss approximation was formulated using CasADi~\citep{casadi} and solved using the primal-dual interior-point NLP solver Fatrop~\citep{fatrop}. The solver was initialized to a random value of the torque distribution ratio. The polynomials used to obtain the results in this section were of the 10th degree.

\begin{figure*}
\centering
\begin{subfigure}{0.49\textwidth}
    \includegraphics[width=\linewidth]{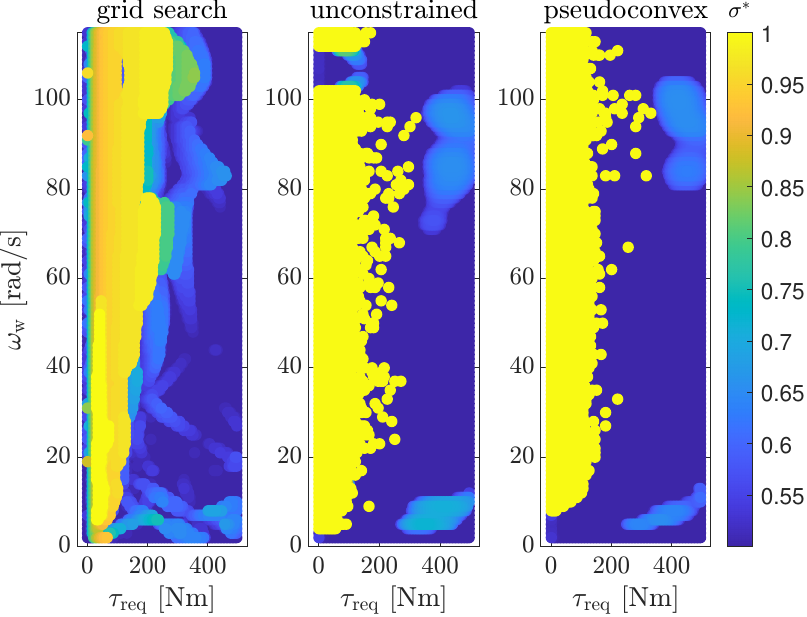}
    \caption{Equal motors}
    \label{fig:split_map_1}
\end{subfigure}
\hfill
\begin{subfigure}{0.49\textwidth}
    \includegraphics[width=\linewidth]{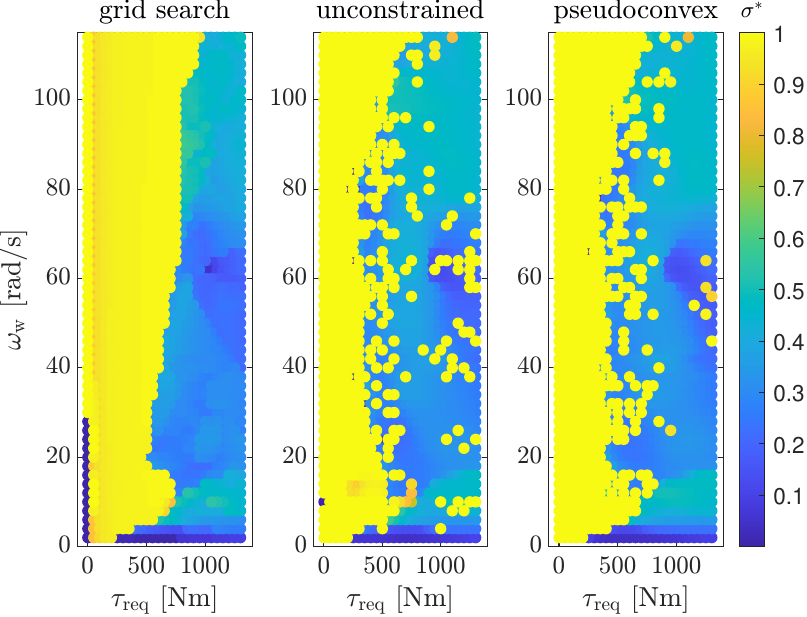}
    \caption{Unequal motors}
    \label{fig:split_map_2}
\end{subfigure}
\caption{Optimal torque distribution ratios obtained with different optimization methods and polynomials of 10th degree}
\label{fig:split_map}
\end{figure*}

Figure~\ref{fig:split_map_1} shows the optimal torque distribution ratios for equal motors with varying torque demand and wheel speed, obtained with different optimization methods. For torque demands higher than what is shown, all methods suggest a distribution ratio of 0.5. The first thing to note are the "smudges" of distribution ratios different than 0.5 for higher torque demands, caused by the flatness of the objective function around that area, resulting in multiple points with almost identical objective function values. Additionally, one can notice "outliers" of high torque distribution ratios for both polynomial approximations, which should be attributed to multiple local minima of the objective function combined with a random initialization of the gradient-based solver. Both effects could be mitigated by additional modifications of the optimization results. With this in mind, the optimal distribution ratios obtained with different methods are similar in the majority of the expected operating range.

The optimal ratios with unequal motors, shown in Figure~\ref{fig:split_map_2}, show that the optimal torque distribution ratios vary more with such a powertrain configuration. The method based on unconstrained polynomials differs the most from the benchmark, while the one based on pseudoconvex polynomials exhibits similar effects as in the case with equal motors. Finally, the results with both configurations were deemed satisfactory for this stage of the research, i.e., without any additional processing of the gradient-based optimization results.

\begin{figure*}
\centering
\begin{subfigure}{0.49\textwidth}
    \includegraphics[width=\linewidth]{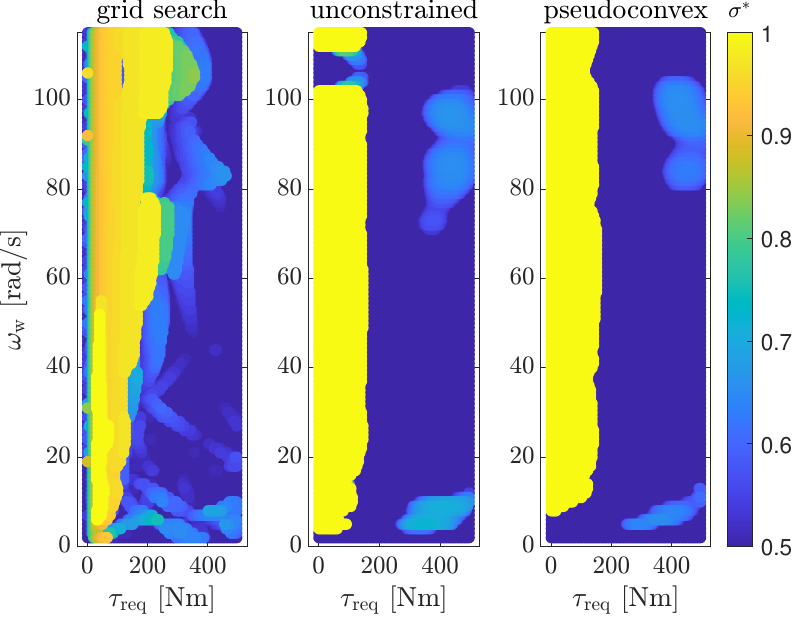}
    \caption{Equal motors}
    \label{fig:split_map_1_global}
\end{subfigure}
\hfill
\begin{subfigure}{0.49\textwidth}
    \includegraphics[width=\linewidth]{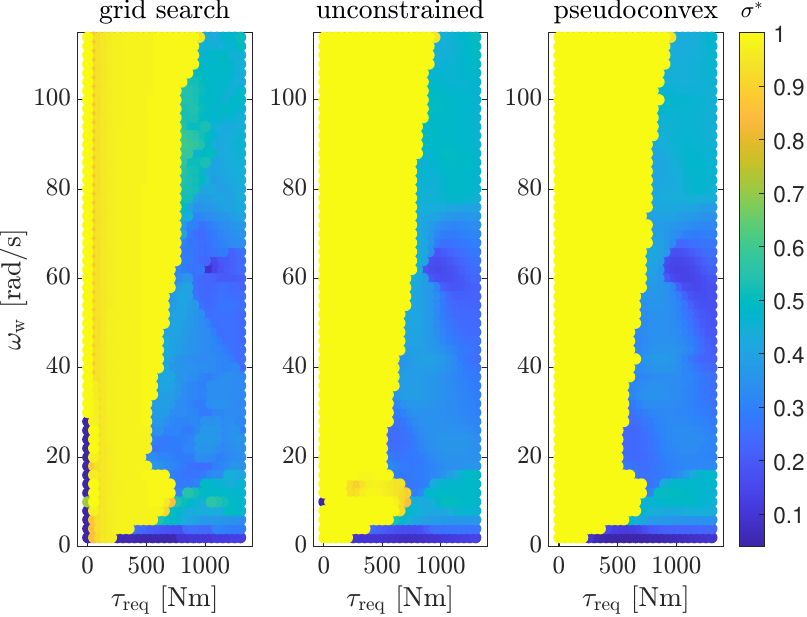}
    \caption{Unequal motors}
    \label{fig:split_map_2_global}
\end{subfigure}
\caption{Optimal torque distribution ratios obtained with different methods and polynomials of 10th degree, when considering the KKT conditions}
\label{fig:split_map_global}
\end{figure*}

However, if only the total powertrain losses are considered in the objective function and are represented as polynomials, using a method based on the KKT conditions might be beneficial. By considering the roots of the objective function derivative, it is possible to find the global optimum. Figure \ref{fig:split_map_global} shows the optimal torque distribution ratios obtained with this method. Although the case with equal motors is not identical to the benchmark, for unequal motors (i.e., a more challenging torque allocation problem) the deviation is significantly reduced.

\subsection{Driving cycle experiments}
\begin{figure}
    \centering
    \includegraphics[width=\linewidth]{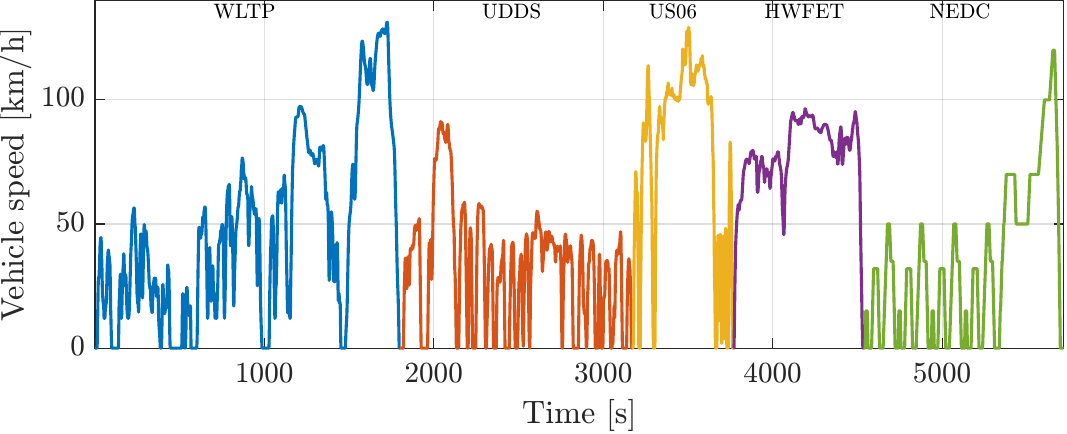}
    \caption{Desired vehicle speed profiles for the different driving cycles}
    \label{fig:cycles}
\end{figure}

Here we present the results of minimizing electric energy consumption obtained on several standard driving cycles. The cycles are defined by their vehicle speed profiles, as shown in Figure~\ref{fig:cycles}. The sampling time is set to $T_s=1$ s.

The benchmark method was the one presented in~\citep{skugor2024vsd}, based on a grid search and denoted by GS. The method based on loss approximation with unconstrained polynomials is denoted by UP, while the one based on pseudoconvex polynomials is denoted by PP. Additionally, the methods based on KKT conditions for the two variants of polynomials are denoted by the extension KKT. Total torque demand for one side of the vehicle is calculated based on the speed profile of the driving cycle, using a discrete version of the vehicle dynamics model presented in Section~\ref{sec:model}.

Tables~\ref{tab:energy_equal} and~\ref{tab:energy_unequal} show the results obtained with different powertrain configurations, torque allocation methods and driving cycles. In all methods except GS, polynomials of 10th degree were used. Results with equal motors suggest that there is a slight advantage to using unconstrained polynomials with equal motors. However, with unequal motors (a more complicated torque allocation problem), the approach based on pseudoconvex polynomials leads to a smaller increase in energy consumption. Interestingly, the methods based on KKT conditions produce similar results with equal motors. However, in the case of unequal motors, a clear advantage of such an approach can be seen, significantly reducing the energy consumption when compared to other methods. Finally, both gradient-based methods indicate an average energy consumption increase of 1.5\% (with a maximum of 2.37\%), which would lead to a relatively small reduction in EV driving range compared to other factors such as environment temperature, battery state of health etc.

\begin{table}
\centering
\caption{Total energy consumption (in kWh) on certification driving cycles with equal motors}
\label{tab:energy_equal}
\begin{tabular}{cccccc}
\hline\noalign{\smallskip}
method              & WLTP & UDDS & US06 & HWFET & NEDC \\
\noalign{\smallskip}\hline\noalign{\smallskip}
GS & 2.7613 & 1.0337 & 1.8604 & 1.9525 & 1.128 \\ \noalign{\smallskip}
\multirow[t]{2}{*}{UP} & 2.8093 & 1.0506 & 1.8744 & 1.9884 & 1.139 \\ 
                    & (+1.74\%) & (+1.63\%) & (+0.75\%) & (+1.84\%) & (+0.97\%) \\ \noalign{\smallskip}
\multirow[t]{2}{*}{PP} & 2.8103 & 1.0582 & 1.8936 & 1.9883 & 1.1462 \\
                    & (+1.77\%) & (+2.37\%) & (+1.79\%) & (+1.83\%) & (+1.61\%) \\ \noalign{\smallskip}
\multirow[t]{2}{*}{UP-KKT} & 2.813 & 1.0506 & 1.8796 & 1.9898 & 1.1388 \\ 
                    & (+1.87\%) & (+1.64\%) & (+1.03\%) & (+1.91\%) & (+0.96\%) \\ \noalign{\smallskip}
\multirow[t]{2}{*}{PP-KKT} & 2.81311 & 1.058 & 1.8967 & 1.9901 & 1.146 \\
                    & (+1.87\%) & (+2.35\%) & (+1.95\%) & (+1.92\%) & (+1.60\%) \\
\noalign{\smallskip}\hline
\end{tabular}
\end{table}

\begin{table}
\centering
\caption{Total energy consumption (in kWh) on certification driving cycles with unequal motors}
\label{tab:energy_unequal}
\begin{tabular}{cccccc}
\hline\noalign{\smallskip}
method              & WLTP & UDDS & US06 & HWFET & NEDC \\
\noalign{\smallskip}\hline\noalign{\smallskip}
GS & 3.2546 & 1.3023 & 2.1746 & 2.2201 & 1.3325 \\ \noalign{\smallskip}
\multirow[t]{2}{*}{UP} & 3.3043 & 1.3235 & 2.2098 & 2.2429 & 1.3466 \\ 
                    & (+1.53\%) & (+1.63\%) & (+1.62\%) & (+1.03\%) & (+1.06\%) \\ \noalign{\smallskip}
\multirow[t]{2}{*}{PP} & 3.2894 & 1.3229 & 2.2026 & 2.2383 & 1.3399 \\
                    & (+1.07\%) & (+1.59\%) & (+1.29\%) & (+0.82\%) & (+0.56\%) \\ \noalign{\smallskip}
\multirow[t]{2}{*}{UP-KKT} & 3.2709 & 1.3078 & 2.1858 & 2.2333 & 1.336 \\ 
                    & (+0.50\%) & (+0.42\%) & (+0.52\%) & (+0.59\%) & (+0.26\%) \\ \noalign{\smallskip}
\multirow[t]{2}{*}{PP-KKT} & 3.2706 & 1.3079 & 2.1859 & 2.2333 & 1.3355 \\
                    & (+0.49\%) & (+0.41\%) & (+0.52\%) & (+0.59\%) & (+0.22\%) \\
\noalign{\smallskip}\hline
\end{tabular}
\end{table}

\subsection{Computational demands}
One of the main goals of the method is to enable real-time optimization, with possibly varying objective or constraints. Across all the experiments, the mean execution time of the gradient-based methods was around 3 ms, with a maximum execution time of around 15 ms. Although additional experiments on application-specific hardware should be conducted, these execution times suggest that the methods are real-time feasible (with the assumed sampling time of 1 s and an order of magnitude difference in runtime between a PC and embedded hardware). Finally, the execution time can be further reduced by a careful initialization of the solver, departing from the default settings or performing hardware-specific code optimization.

\section{Conclusion}
\label{sec:outro}
In this paper, we have presented a method of approximating electric motor losses with polynomials of arbitrary order and with certain desired properties. To ensure positivity and monotonicity of the loss function approximation, we employed a framework based on sum of squares programming. This enables a more realistic approximation of motor losses even with very sparse or imperfect datasets, while retaining computational efficiency.

We also proposed a gradient-based torque allocation strategy for straight-line driving of an electric vehicle with equal or unequal motors. With the motor losses approximated by (pseudoconvex) polynomials, the nonlinear optimization problem can be solved in real-time, enabling a more flexible approach (e.g., when handling different operational constraints) and easier integration with other vehicle control systems when compared with offline-optimized and rule-based solutions. Additionally, we demonstrated that a method based on the KKT conditions can avoid suboptimal solutions caused by local minima.

\section*{Declarations}
\textbf{Ethics approval and consent to participate:}
Not applicable.

\vspace{.2cm}\noindent
\textbf{Consent for publication:}
Not applicable.

\vspace{.2cm}\noindent
\textbf{Funding:} J.K.H. and Š.I. have been supported in part by the Croatian Science Foundation under the project "PVDC - Predictive vehicle dynamics control", UIP-2019-04-6487, and the European union through NextGenerationEU. Š.I. has also been supported in part by the European union through the European Regional Development Fund via Operative Programme Competence and Cohesion 2014-2020 for Croatia within the project EVBattPredtect, contract no. KK.01.1.1.07.0029.

\vspace{.2cm}\noindent
\textbf{Availability of data and materials:}
Upon request.

\vspace{.2cm}\noindent
\textbf{Competing interests:}
The authors declare that they have no competing interests.

\vspace{.2cm}\noindent
\textbf{Authors' contributions:}
Methodology: J.K.H., Š.I., B.Š., J.D.; software: J.K.H., Š.I., B.Š.; visualization: J.K.H.; analysis and discussion of methods and results: J.K.H., Š.I., B.Š., J.D.;  All authors read and approved the final manuscript.

\vspace{.2cm}\noindent
\textbf{Acknowledgements:}
Not applicable.

\bibliographystyle{class/spbasic}      % basic style, author-year citations
\bibliography{content/bibliography}   % name your BibTeX data base

@article{eff_tv_survey,
    author = {Sforza, Andrea and Lenzo, Basilio and Timpone, Francesco},
    year = {2019},
    month = {12},
    pages = {},
    title = {A state-of-the-art review on torque distribution strategies aimed at enhancing energy efficiency for fully electric vehicles with independently actuated drivetrains},
    journal = {International Journal of Mechanics and Control}
}

@ARTICLE{eff_tv_nmpc,
    author={Parra, Alberto and Tavernini, Davide and Gruber, Patrick and Sorniotti, Aldo and Zubizarreta, Asier and Pérez, Joshué},
    journal={IEEE Transactions on Vehicular Technology}, 
    title={On Nonlinear Model Predictive Control for Energy-Efficient Torque-Vectoring}, 
    year={2021},
    volume={70},
    number={1},
    pages={173-188},
    doi={https://doi.org/10.1109/TVT.2020.3022022}
}

@book{boyd2004convex,
  title={Convex optimization},
  author={Boyd, Stephen and Vandenberghe, Lieven},
  year={2004},
  publisher={Cambridge university press}
}

@ARTICLE{dizqah2016,
  author={Dizqah, Arash M. and Lenzo, Basilio and Sorniotti, Aldo and Gruber, Patrick and Fallah, Saber and De Smet, Jasper},
  journal={IEEE Transactions on Industrial Electronics}, 
  title={A Fast and Parametric Torque Distribution Strategy for Four-Wheel-Drive Energy-Efficient Electric Vehicles}, 
  year={2016},
  volume={63},
  number={7},
  pages={4367-4376},
  doi={https://doi.org/10.1109/TIE.2016.2540584}
}

@ARTICLE{filippis2018,
  author={De Filippis, Giovanni and Lenzo, Basilio and Sorniotti, Aldo and Gruber, Patrick and De Nijs, Wouter},
  journal={IEEE Transactions on Vehicular Technology}, 
  title={Energy-Efficient Torque-Vectoring Control of Electric Vehicles With Multiple Drivetrains}, 
  year={2018},
  volume={67},
  number={6},
  pages={4702-4715},
  doi={https://doi.org/10.1109/TVT.2018.2808186}
}

@INPROCEEDINGS{menner2021,
  author={Menner, Marcel and Di Cairano, Stefano},
  booktitle={2021 IEEE Vehicle Power and Propulsion Conference (VPPC)}, 
  title={Kernel Regression for Energy-Optimal Control of Fully Electric Vehicles}, 
  year={2021},
  volume={},
  number={},
  pages={1-6},
  doi={https://doi.org/10.1109/VPPC53923.2021.9699361}
}

@book{parrilo2012,
    author = {Blekheran, Grigoriy and Parrilo, Pablo A. and Thomas, Rekha R.},
    title = {Semidefinite Optimization and Convex Algebraic Geometry},
    publisher = {Society for Industrial and Applied Mathematics},
    year = {2012},
    doi = {https://doi.org/10.1137/1.9781611972290},
    edition   = {}
}

@article{skugor2024vsd,
    author = {Branimir Škugor and Joško Deur and Weitian Chen and Yijing Zhang and Edward Dai},
    title = {A parameter-optimised rule-based control strategy for front-rear torque vectoring in electric vehicles with multiple motors and disconnect clutches},
    journal = {Vehicle System Dynamics},
    pages = {1--27},
    year = {2024},
    publisher = {Taylor \& Francis},
    doi = {https://doi.org/10.1080/00423114.2024.2430581},
}

@article{lenzo2017torque,
    author = {Lenzo, B. and De Filippis, G. and Dizqah, A. M. and Sorniotti, A. and Gruber, P. and Fallah, S. and De Nijs, W.},
    title = {Torque Distribution Strategies for Energy-Efficient Electric Vehicles With Multiple Drivetrains},
    journal = {Journal of Dynamic Systems, Measurement, and Control},
    volume = {139},
    number = {12},
    pages = {121004},
    year = {2017},
    month = {08},
    issn = {0022-0434},
    doi = {https://doi.org/10.1115/1.4037003},
}

@INPROCEEDINGS{magnani2005tractable,
  author={Magnani, A. and Lall, S. and Boyd, S.},
  booktitle={Proceedings of the 44th IEEE Conference on Decision and Control}, 
  title={Tractable fitting with convex polynomials via sum-of-squares}, 
  year={2005},
  volume={},
  number={},
  pages={1672-1677},
  doi={https://doi.org/10.1109/CDC.2005.1582399}
}

@inproceedings{fatrop,
  title={Fatrop: A fast constrained optimal control problem solver for robot trajectory optimization and control},
  author={Vanroye, Lander and Sathya, Ajay and De Schutter, Joris and Decr{\'e}, Wilm},
  booktitle={2023 IEEE/RSJ International Conference on Intelligent Robots and Systems (IROS)},
  pages={10036--10043},
  year={2023},
  organization={IEEE}
}

@inproceedings{yalmip,
address = {Taipei, Taiwan},
author = {L{\"{o}}fberg, J.},
booktitle = {In Proceedings of the CACSD Conference},
title = {{YALMIP} : A Toolbox for Modeling and Optimization in {MATLAB}},
year = {2004}
}

@article{yalmip_sos,
author = {L{\"{o}}fberg, Johan},
journal = {IEEE Transactions on Automatic Control},
number = {5},
pages = {1007--1011},
title = {Pre- and post-processing sum-of-squares programs in practice},
volume = {54},
year = {2009}
}

@Article{casadi,
  author = {Joel A E Andersson and Joris Gillis and Greg Horn
            and James B Rawlings and Moritz Diehl},
  title = {{CasADi} -- {A} software framework for nonlinear optimization
           and optimal control},
  journal = {Mathematical Programming Computation},
  volume = {11},
  number = {1},
  pages = {1--36},
  year = {2019},
  publisher = {Springer},
  doi = {https://doi.org/10.1007/s12532-018-0139-4}
}

@article{skugor2024optimization,
author={{\v{S}}kugor, Branimir
and Deur, Jo{\v{s}}ko
and Chen, Weitian
and Zhang, Yijing
and Dai, Edward},
title={Optimization of straight-line driving torque vectoring for energy-efficient operation of electric vehicles with multiple motors and disconnect clutches},
journal={Optimization and Engineering},
year={2024},
month={Jul},
day={17},
issn={1573-2924},
doi={https://doi.org/10.1007/s11081-024-09902-7}
}

@misc{mosek,
  title={{MOSEK} optimization suite},
  author={{MOSEK ApS}},
  year={2017}
}

@article{curmei2023shape,
  title={Shape-constrained regression using sum of squares polynomials},
  author={Curmei, Mihaela and Hall, Georgina},
  journal={Operations Research},
  volume={73},
  number={1},
  pages={543--559},
  year={2023},
  publisher={INFORMS},
  doi={https://doi.org/10.1287/opre.2021.0383}
}

@article{Hilbert1888,
author={Hilbert, David},
title={Über die Darstellung definiter Formen als Summe von Formenquadraten},
journal={Mathematische Annalen},
year={1888},
month={Sep},
day={01},
volume={32},
number={3},
pages={342-350},
issn={1432-1807},
doi={https://doi.org/10.1007/BF01443605},
}

@inproceedings{deur2012modeling,
  title={Modeling and low-level control of range extended electric vehicle dynamics},
  author={Deur, Jo{\v{s}}ko and Cipek, Mihael and {\v{S}}kugor, Branimir and Petri{\'c}, Jo{\v{s}}ko},
  booktitle={1st Biannual International Conference on Powertrain Modelling and Control (PMC2012)},
  year={2012}
}

@article{miller2011gm,
  title={The GM “Voltec” 4ET50 multi-mode electric transaxle},
  author={Miller, Michael A and Holmes, Alan G and Conlon, Brendan M and Savagian, Peter J},
  journal={SAE International Journal of Engines},
  volume={4},
  number={1},
  pages={1102--1114},
  year={2011},
  publisher={JSTOR}
}

@inproceedings{prost2024energy,
  title={Energy-Efficient Optimal Torque Vectoring for a Four-Motor High-Performance Electric Vehicle},
  author={Prost, Matt{\'e}o and Cvok, Ivan and Velenis, Efstathios},
  booktitle={Advanced Vehicle Control Symposium},
  pages={804--811},
  year={2024},
  organization={Springer}
}

@article{mangasarian,
author = {Mangasarian, O. L.},
title = {Pseudo-Convex Functions},
journal = {Journal of the Society for Industrial and Applied Mathematics Series A Control},
volume = {3},
number = {2},
pages = {281-290},
year = {1965},
doi = {https://doi.org/10.1137/0303020}
}

@article{chen2011fast,
  title={Fast and global optimal energy-efficient control allocation with applications to over-actuated electric ground vehicles},
  author={Chen, Yan and Wang, Junmin},
  journal={IEEE Transactions on Control Systems Technology},
  volume={20},
  number={5},
  pages={1202--1211},
  year={2011},
  publisher={IEEE}
}

@article{dizqah2020,
title = {A Non-Convex Control Allocation Strategy as Energy-Efficient Torque Distributors for On-Road and Off-Road Vehicles},
journal = {Control Engineering Practice},
volume = {95},
pages = {104256},
year = {2020},
issn = {0967-0661},
doi = {https://doi.org/10.1016/j.conengprac.2019.104256},
author = {A.M. Dizqah and B.L. Ballard and M.V. Blundell and S. Kanarachos and M.S. Innocente},
}

@article{ou2020,
doi = {https://doi.org/10.1088/1742-6596/1550/4/042020},
year = {2020},
publisher = {IOP Publishing},
volume = {1550},
number = {4},
author = {Ou, Yang and Wang, Peng and Xu, Lei and Fan, Jie and Zhou, Zhou and Li, Zhe and Bai, Qin and Zhang, Yuanwei and Gao, Zihao},
title = {Torque allocation strategy for two axles four wheel drive electric vehicle with improvement of economy and stability},
journal = {Journal of Physics: Conference Series}
}

@ARTICLE{koehler2017,
  author={Koehler, Stefan and Viehl, Alexander and Bringmann, Oliver and Rosenstiel, Wolfgang},
  journal={IEEE Intelligent Transportation Systems Magazine}, 
  title={Energy-Efficiency Optimization of Torque Vectoring Control for Battery Electric Vehicles}, 
  year={2017},
  volume={9},
  number={3},
  pages={59-74},
  doi={https://doi.org/10.1109/MITS.2017.2709799}
}

@book{biegler2010nonlinear,
  title={Nonlinear programming: concepts, algorithms, and applications to chemical processes},
  author={Biegler, Lorenz T},
  year={2010},
  publisher={SIAM}
}

@ARTICLE{xu2020,
  author={Xu, Ying and Jiang, Lunyao and Wei, Bo and Qiu, Li},
  journal={IEEE Access}, 
  title={An Optimal Torque Distribution Strategy for Four-Motorized-Wheel Electric Vehicle Considering Energy Conversation}, 
  year={2020},
  volume={8},
  number={},
  pages={135975-135988},
  doi={https://doi.org/10.1109/ACCESS.2020.3008068}
}

@article{yang2017research,
  title={Research on Optimized Torque-Distribution Control Method for Front/Rear Axle Electric Wheel Loader},
  author={Yang, Zhiyu and Wang, Jixin and Gao, Guangzong and Shi, Xiangyun},
  journal={Mathematical Problems in Engineering},
  volume={2017},
  number={1},
  pages={7076583},
  year={2017},
  publisher={Wiley Online Library},
  doi={https://doi.org/10.1155/2017/7076583}
}

@Article{grebe2011,
author={Grebe , Uwe D.
and Nitz, Larry T.},
title={Voltec -- The Propulsion System for Chevrolet Volt and Opel Ampera},
journal={MTZ worldwide eMagazine},
year={2011},
month={May},
day={01},
volume={72},
number={5},
pages={4-11},
issn={2192-9114},
doi={https://doi.org/10.1365/s38313-011-0046-9}
}

\end{document}